\documentclass{iopart}

\usepackage[utf8x]{inputenc}
\usepackage{iopams}
\usepackage{subfigure,braket,color}
\usepackage[english]{babel}
\usepackage{graphicx}
\usepackage{sistyle}

\newcommand{\Z}{\mathbb{Z}}
\newcommand{\cm}[1]{[ #1 ]}
\newcommand{\acm}[1]{\left\{ #1\right \}}

\newcommand{\trace}{\mathrm{Tr}}
\newcommand{\Var}{\mathrm{Var}}

\renewcommand{\d}{\!\mathrm{d}}
\newcommand{\mean}[1]{E\left[ #1 \right] }
\newcommand{\abs}[1]{\left|#1\right|}

\newcommand{\normalorder}[1]{\colon #1 \colon}

\newcommand{\heff}{h_{\text{eff}}}
\newcommand{\tildeheff}{\tilde h_{\text{eff}}}

\newenvironment{align}{\begin{eqnarray}}{\end{eqnarray}}
\newenvironment{align*}{\begin{eqnarray*}}{\end{eqnarray*}}
\newenvironment{multline}{\begin{eqnarray}}{\end{eqnarray}}
\newenvironment{multline*}{\begin{eqnarray*}}{\end{eqnarray*}}
\newenvironment{widetext}{}{}

\begin{document}
\title{Phase transitions in an Ising chain interacting with a single mode cavity field}
\author{S{\o}ren Gammelmark, Klaus M{\o}lmer}
\address{Lundbeck Foundation Theoretical Center for
  Quantum System Research, Department of Physics and Astronomy,
  University of Aarhus, DK 8000 Aarhus C, Denmark}
\ead{gammelmark@phys.au.dk}

\begin{abstract}
We investigate the thermodynamics of a combined Dicke- and Ising-model which exhibits a rich phenomenology arising from the second order and quantum phase transitions from the respective models. The partition function is calculated using mean field theory, and the free energy is analyzed in detail to determine the complete phase diagram for the system. The analysis reveals both first- and second-order Dicke phase transitions into a super-radiant state, and the cavity mean-field in this regime acts as an effective magnetic field, which restricts the Ising chain dynamics to parameter ranges away from the Ising phase transition. Physical systems with a first order phase transitions are natural candidates for metrology and calibration purposes, and we apply filter theory to show that the sensitivity of the physical system to temperature and external fields reaches the $1/N$ Heisenberg limit. \end{abstract}

\section{Introduction}

The understanding of the remarkable and useful properties of matter in different phases and of the critical behaviour near phase transitions presents ongoing challenges in theoretical and experimental physics \cite{Coldea2010,Bloch_2002,vidal_class_2008,verstraete_criticality_2006,zhou_signature_2010}. Entirely new types of phase transition phenomena become relevant with the ability to control and engineer microscopic interactions and systems, e.g., in cold atom experiments with spinor Bose-Einstein condensates and with Fermi gases \cite{bloch_many-body_2008, ueda_spinor_2010}. Indeed, a whole branch of quantum computing research attempts to use candidate systems for quantum computing as quantum simulators in which quantum gates are operated so as to simulate suitable inter-particle interactions \cite{lewenstein_ultracold_2007} and in this way implement theoretical phase transition models in a quantum analog computer. \par

Local interactions between nearest neighbours in a spin chain lead to Ising-, Heisenberg-, and other interaction models with phase transitions at definite values of the interaction strengths and external controllable parameters, such as a bias magnetic field. While these interactions are reliable models of, e.g., magnetic interactions in solids, they can also be engineered exactly among trapped atoms or ions, with the added experimental possibility to control the sign and magnitude of the interactions with laser beams and the spin temperature by optical pumping \cite{porras_effective_2004}.

Atomic and optical systems also permit the engineering of interactions between a large number of atoms and a quantized oscillator mode. Such systems are implemented in various schemes for quantum computing, where the oscillator mode is used as a data bus between the atomic quantum bits. Beyond a critical coupling strength to the oscillator and below a critical temperature the system undergoes a phase transition, and the thermodynamic ground state acquires a macroscopic excitation of the oscillator mode. This phase-transition was first discussed by Dicke \cite{dicke_coherence_1954}, and the Dicke phase transition has since then been studied extensively \cite{wang_phase_1973,hepp_equilibrium_1973}, and was recently observed in experiments with cold atoms in an optical cavity \cite{baumann_dicke_2010} using techniques similar to those described in \cite{dimer_proposed_2007}.

It is conceivable that one can implement both the Ising and the Dicke interaction Hamiltonians in many different ways using atoms and cavities, nano-mechanical devices or collective vibrational motion of the atoms or ions. The partition function of such a combined Dicke-Ising model was  determined recently in the thermodynamic limit \cite{lee_first-order_2004}, and in this article we study the phase diagram and further properties of the system and possible applications.

First order phase transitions represent discontinuous changes and hence a high sensitivity to variations in the values of the physical parameters of the model near the critical point. Sensitivity of quantum systems in fundamental metrology is a very active research field where much research has been devoted to identify how the sensitivity depends on the number of particles $N$. The "standard limit" $1/\sqrt{N}$, is replaced by the "Heisenberg limit" $1/N$ within different realizations of non-interacting particles \cite{wineland_squeezed_1994, giovannetti_quantum_2006,caves_quantum-mechanical_1981,braunstein_statistical_1994}, while more rapid decrease with particle number has been proposed in different models of interacting particles \cite{negretti_quantum_2008,boixo_generalized_2007,pezze_entanglement_2009}. In this manuscript we argue for a power law decrease similar to the Heisenberg limit for measurements of temperature with our interacting system.  

The manuscript is organized as follows: In section \ref{sec:DickeIsingModel} we outline the model and discuss the two limiting Ising and Dicke regimes and their phenomenology. In section \ref{sec:PartitionFunction} we review the phase diagram calculation \cite{lee_first-order_2004} and we recast this calculation in terms of mean-field theory. In section \ref{sec:Phasediagram} we discuss the phase diagram in detail. In Section V we analyze the application of the phase transition for metrology. Finally, we conclude in section \ref{sec:Conclusion}.

\section{The Dicke-Ising model} \label{sec:DickeIsingModel}

\subsection{The Ising Model}

Consider a one-dimensional chain of $N$ spins or two-level atoms realizing an Ising model with a transverse magnetic field,
\begin{align}
 H_\text{Ising} &= -h \sum_{i=1}^N \sigma_i^z - J\sum_i \sigma_i^y \sigma_{i+1}^y \label{eq:HIsingSpin}
\end{align}
where $h \geq 0$ is the transverse field and $J \geq 0$ is the interaction strength between neighbouring spins and $\sigma_{N+1}^y \equiv \sigma_1^y$.

The model is not trivially easy to diagonalize, as the transverse field operators $\sigma_i^z$ do not commute with the interaction terms $\sigma_i^y\sigma_{i+1}^y$. In general, for $h \gg J$ the transverse field dominates in which case we expect $\braket{\sigma_i^y} \approx 0$ and $\braket{\sigma_i^z} = 1$. For a strong coupling $J \gg h$ the interaction term favours parallel spin in the $y$-direction so $\braket{\sigma_i^z} \approx 0$ and the system possesses two states of equal energy with $\braket{\sigma_i^y} = + 1$, or $\braket{\sigma_i^y} = -1$.

An (almost) exact diagonalization of the Ising Hamiltonian can be performed \cite{pfeuty_one-dimensional_1970,lieb_two_1961} using a Jordan-Wigner transformation which maps the spin-operators to fermionic operators. The particular mapping we employ here is $c_n = i\prod_{j < n} \sigma_j^z \sigma_n^\dagger$, where $\sigma_n^\dagger = (\sigma_n^x - i\sigma_n^y)/2$ is the $n$'th spin-$z$ step-up operator. Multiplying the spin-operators with phase factors, $\sigma_j^z$, depending on the spins at previous locations in the spin-chain leads to the fermionic anti-commutator relations $\acm{c_n, c_{n'}^\dagger} = \delta_{n,n'}$. The fermionic number operator is $c_n^\dagger c_n = \sigma_n \sigma_n^\dagger = (1 - \sigma_n^z)/2$ and the state with all spins pointing in the $z$-direction is mapped to the fermionic vacuum while $c_n^\dagger$ flips a spin from up to down thereby creating a fermion. The inverse mapping reads $\sigma_n^\dagger = -i\prod_{j<n} (1 - 2c_j^\dagger c_n) c_n$ such that the $y$-$y$-interaction term in the Ising Hamiltonian becomes $-J\sigma_i^y \sigma_{i+1}^y = -J(c_i^\dagger - c_i)(c_{i+1}^\dagger + c_{i+1})$ for $i < N$. Due to the periodic boundary conditions we cannot remove the intermediate $(1-2c_j^\dagger c_j)$-factors in the final $y$-$y$-term  $-J\sigma_N^y \sigma_1^y$. This term can be written as $J (c_N^\dagger - c_N)(c_1^\dagger + c_1) \prod_{j=1}^N (1 - 2 c_j^\dagger c_j) $ and by adding and subtracting $J(c_N^\dagger - c_N)(c_1^\dagger + c_1)$ we can write the resulting Hamiltonian as
\begin{multline*}
 H_\text{Ising} = -h \sum_i (1 - 2c_i^\dagger c_i) - J \sum_i c_i^\dagger c_{i+1} + c_{i+1}^\dagger c_i + c_{i+1} c_i + c_i^\dagger c_{i+1}^\dagger \\
  + J(c_N^\dagger - c_N)(c_1^\dagger + c_1) (1 + \prod_{j=1}^N (1 - 2 c_j^\dagger c_j))
\end{multline*}
where the last term is of relative order $1/N$ since $1 + \prod_j (1 - 2 c_j^\dagger c_j)$ is a projection operator onto the subspace of even total spin in the $z$-direction and therefore is of order 1.

By neglecting this less important final term the Hamiltonian is quadratic in the fermionic operators, and can be diagonalized using a Bogoliubov-transformation. The Bogoliubov-transformation can be decomposed into a single-particle basis change to momentum space and quasi-particle operators connecting particles of opposite momenta \cite{pfeuty_one-dimensional_1970,lieb_two_1961,sachdev_quantum_2001} and we get:
\begin{align}
  H_\text{Ising} = \sum_{k \in \Z_N} \epsilon(k) \left(\gamma_k^\dagger \gamma_k - \frac{1}{2}\right) + O(1/N), \label{eq:HIsingDiagonal}
\end{align}
where the operators $\gamma_k$ are the Bogoliubov transformed fermionic operators, $\epsilon(k) = 2(J^2 + h^2 - 2 J h \cos(k))^{1/2}$ and $k \in \Z_N$ has the form $2\pi n/N$ and is a reciprocal lattice-vector to the chain-lattice. The eigenmodes of $H_\text{Ising}$ correspond to free fermions with a dispersion relation given by $\epsilon(k)$. Note that (\ref{eq:HIsingDiagonal}) has a single unique ground state in contrast to (\ref{eq:HIsingSpin}) which has a two-fold degeneracy. This difference is due to the omitted term and will not be relevant in the thermodynamic limit.

Since the spectrum of the Ising model is so simple we can calculate the partition function for the system. The partition function is given by $Z^0_\text{Ising} = \trace(\exp(-\beta H_\text{Ising}))$ where $\beta$ is the inverse temperature. We can easily calculate this trace,
\begin{align*}
 Z^0_\text{Ising}(\beta, h, J) &= \prod_{k\in\Z_N} \trace\left[\exp(-\beta \epsilon(k) (\gamma_k^\dagger \gamma_k - 1/2))\right] \\
  &= \prod_{k\in\Z_N} 2\cosh(\beta \epsilon(k)/2).
\end{align*}
Using this result we can also calculate the free energy given by $F_\text{Ising} = -\beta^{-1} \log Z^0_\text{Ising}$. The diagonalization of $H_\text{Ising}$ is exact to order $1/N$ and to the same precision we can replace the sum over $k \in \Z_N$ by an integral,
\begin{align}
 F_\text{Ising}(\beta, h, J) &= -\frac{1}{\beta}\sum_{k\in\Z_N} \log(2\cosh(\beta\epsilon(k)/2)) \nonumber \\
  &\approx -N \int_{-\pi}^\pi \frac{\d k}{2\pi\beta} \log(2\cosh(\beta\epsilon(k, h, J))). \label{eq:IsingFreeEnergy}
\end{align}
Note that $\beta \epsilon(k, h, J)/2 = \beta J (1 + (h/J)^2 - 2(h/J)\cos(k))^{1/2}$, and hence it is convenient to parametrize $F_\text{Ising}$ with the dimensionless quantities $\tilde\beta = \beta J$ and $\tilde h = h/J$ in the following.

The Ising model exhibits an infinite order quantum phase-transition between a paramagnetic $h \gg J$ and ferromagnetic $J \gg h$ phase with critical point at $J = h$ where a non-analyticity arises in $\epsilon(k)$ at $k = 0$. This system has been studied extensively \cite{sachdev_quantum_2001} and is a prime example of a quantum phase transition. The quantum phase-transition does not survive to finite temperatures, but it has important consequences for the finite-temperature behaviour of the system near the critical point.

\subsection{The Dicke model}
Another phase transition model consists of spins or two-level atoms coupled to a harmonic oscillator mode with frequency $\omega$,
\begin{align}
 H_\text{osc} &= \omega a^\dagger a, \label{eq:H-cavity}
\end{align}
through the interaction
\begin{align}
 V &= \frac{g}{\sqrt{N}} \sum_i \sigma_i^x (a + a^\dagger), \label{eq:V-cavity}
\end{align}
where $\sigma^x_i$ is the Pauli $x$-matrix acting on the $i$'th spin and $a$ is the harmonic oscillator step down operator. We denote the coupling-strength of the oscillator to a single spin by $g/\sqrt{N}$. For a physical implementation with atoms inside a cavity with a mode volume $V$, the quantum field strength per photon is proportional to $1/\sqrt{V}$, and our scaling thus corresponds to atoms with a constant spatial density which is well defined, also in the thermodynamic limit $N\to\infty$.

The explicit form of the interaction is well established in quantum optical systems, and occurs both for two-level systems and for, e.g., Raman processes between two states via an intermediate excited state through absorption and stimulated emission of the quantum field and a classical control field.  We note that in the Jaynes-Cummings model of a single two-level atom and a single field mode,
the rotating wave approximation retains only the terms $\sigma^\dagger a + \sigma a^\dagger$ in $V$ and provides a considerable simplification of the problem. The partition function for the Dicke-model has been calculated analytically in the thermodynamic limit and a second order phase transition has been identified \cite{wang_phase_1973}. While that calculation pertained to the rotating wave approximation it has been shown \cite{hepp_equilibrium_1973} that this does not change the Dicke phase-transition qualitatively. Within our application of a mean field approximation to the combined Dicke-Ising model we shall retain the full interaction (\ref{eq:V-cavity}) as, the rotating wave approximation is more difficult to deal with.

The Dicke model $H_\text{osc}+V$ can be realized experimentally as described in \cite{dimer_proposed_2007} where a dynamical version of the standard Dicke model is investigated in a cavity using four-level atoms coupled by Raman channels. The model parameters of this system are functions of the atomic and field parameters applied and can be tuned over large ranges.

\section{The partition function for the Dicke-Ising model} \label{sec:PartitionFunction}
The combined Dicke-Ising model has total Hamiltonian $H=H_\text{Ising}+H_\text{osc}+V$ with three parameters $g, \ h$, and $J$ which can be varied independently in atomic simulators of the model. With $J = 0$ the Hamiltonian realizes the Dicke model of $N$ two-level atoms interacting with a cavity-field via the dipole interaction. In this regime $2 h$ correspond to the energy-splitting of the individual two-level atoms and $g$ is the coupling strength to the cavity-field.

\subsection{Coherent state integral} \label{sec:WangPartitionFunction}

We will proceed by writing first the expression of the partition function for the full problem $Z = \trace(\exp(-\beta H))$ with $H=H_\text{Ising}+H_\text{osc}+V$.
Following \cite{wang_phase_1973} we write $H$ as
\begin{align*}
    H = \sum_{i=1}^N \left[- h\sigma_i^z - J\sigma_i^y \sigma_{i+1}^y 
     + g \sigma_i^x \left(\frac{a}{\sqrt{N}} + \frac{a^\dagger}{\sqrt{N}}\right) + \omega \frac{a^\dagger}{\sqrt{N}} \frac{a}{\sqrt{N}}  \right].
\end{align*}
We will evaluate the trace over the oscillator-mode in the partition function using the coherent state representation of the field, and for this purpose we should bring $\exp(-\beta H)$ on normal ordered form with respect to the scaled operators $b = a/\sqrt{N}$ and $b^\dagger = a^\dagger / \sqrt{N}$. Due to the scaling, their commutator is $\cm{b, b^\dagger} = 1/N$ and vanishes for $N \to \infty$.

The approximation of neglecting the commutator in the thermodynamic limit can be illuminated by Wick's theorem \cite{fetter_quantum_2003}: Any string of creation- and annihilation operators can be written as their normally ordered form plus additional terms. These terms are given by means of a \emph{contraction} defined by $C(AB) = AB - \normalorder{AB}$, where $\normalorder{AB}$ is the normal ordering of $AB$. Using this notation, Wick's theorem states that
\begin{multline*}
 ABC\ldots Q = \normalorder{ABC\ldots Q} + \sum\normalorder{\text{one contraction}} \\
  + \sum\normalorder{\text{two contractions}} + \ldots
\end{multline*}
Now, in our case $C(b b^\dagger) = b b^\dagger - b^\dagger b = \cm{b, b^\dagger} = 1/N$ and $C(b^\dagger b) = 0$. Hence, applying Wick's theorem to the expansion of the exponential in $Z = \trace(\exp(-\beta H))$, and assuming that the limits in $Z = \lim_{N\to\infty} \lim_{R\to\infty} \sum_{r=0}^R (-\beta H)^r/r!$ can be interchanged, we see that all terms involving contractions will be of order $1/N$ or higher and can therefore be neglected in the thermodynamic limit. More generally any expression of the form $\trace( f(b, b^\dagger) e^{-\beta H})$, where $f$ is a polynomial, can be calculated to an accuracy of $1/N$ using the normal order $\normalorder{f(b, b^\dagger)e^{-\beta H}}$. The validity of this truncation for the Dicke-model was discussed in \cite{hepp_equilibrium_1973}.

When performing the trace of a normally ordered operator it is convenient to use the coherent states, $|\alpha\rangle$, eigenstates of the annihilation operator: $a|\alpha\rangle = \alpha|\alpha\rangle$, which yields
\begin{widetext}
\begin{align*}
 Z = \int\frac{\d^2\alpha}{\pi}
  e^{-\beta \omega \abs{\alpha}^2} \trace_\text{spin}\left(\exp\left(-\beta\sum_i\left[-h\sigma_i^z - J\sigma_i^y \sigma_{i+1}^y + \frac{2g \Re(\alpha)}{\sqrt{N}}\sigma_i^x\right]\right)\right) + O(1/N).
\end{align*}
\end{widetext}
The remaining trace over the spin degrees of freedom is exactly equivalent to the the original Ising model calculation, where the term involving the real part of the complex field argument acts as an additional magnetic field in the $x$ direction, and hence the Ising model is biased by an effective magnetic field in the $xz$-plane with magnitude $\heff^2 = h^2 + 4 g^2 \Re(\alpha)^2/N$. Since this effective field lies within the plane orthogonal to the $y$-axis we can choose the direction of the effective field as a redefined $z$-axis and apply the same diagonalization as in the pure Ising problem.\par

 This yields $Z = \int\frac{\d^2\alpha}{\pi} \exp(-\beta \omega \abs{\alpha}^2 + \log Z^0_\text{Ising}(\beta, \heff, J) )$ and after performing the substitution $z = \alpha/\sqrt{N}$ we get
\begin{multline}
 Z = N\int\frac{\d^2 z}{\pi}
  \exp\left(-N \beta \omega \abs{z}^2 \right) \nonumber \\
  \cdot\exp\left( \frac{N}{2\pi}\int_{-\pi}^\pi \d k \log(2\cosh(\beta \epsilon(k, \heff( \Re(z) ), J)/2)) \right). \label{eq:CoherentStatePartitionFunction}
\end{multline}
where the dispersion relation $\epsilon(k, \heff(x), J)$ is for the effective magnetic field $\heff$ and spin coupling $J$.

\subsection{Mean field theory} \label{sec:MeanField}

We may also attack the original problem with an Ansatz replacing the interaction part $V$ of the Hamiltonian by the mean field expression
\begin{align*}
 V_{MF} &= g N s_x (a + a^\dagger) / \sqrt{N} + 2 g\sum_i \sigma_i^x x - 2 g N s_x x
\end{align*}
where the $c$-number mean fields read $s_x = \braket{\sum_{i=1}^N \sigma_i^x}/N$ and $x = \braket{a + a^\dagger}/2\sqrt{N}$. The last term compensates for double-counting of the interaction energy, while correlations in the mean-field fluctuations $g (\sum_i \sigma_i^x - N s_x) (a + a^\dagger - \sqrt{N} x)/\sqrt{N}$ are omitted.\par

With the Hamiltonian in this form the mean-fields split the Hamiltonian into two separate terms: a classically driven field mode $H_\text{MF,osc} = \omega a^\dagger a + g N s_x(a + a^\dagger)/\sqrt{N}$ and an Ising model with an additional magnetic field component along the $x$-direction, $H_\text{MF,Ising} = -h \sum_i \sigma_i^z + 2g x \sum_i \sigma_i^x - J \sum_i \sigma_i^y \sigma_{i+1}^y$. This mean field Ising Hamiltonian has a field in the $xz$-plane with magnitude $\heff^2 = h^2 + 4 g^2 x^2$. The partition function for the mean field Hamiltonian therefore factors $Z = Z_\text{MF,osc} Z_\text{MF,Ising} \exp(2 \beta g N s_x x)$ and can be readily determined for arbitrary values of the mean field amplitudes $x$ and $s_x$. The corresponding free energy reads
\begin{multline*}
 F = \frac{1}{\beta} \log(1 - e^{-\beta\omega}) - N \frac{g^2 (s_x)^2}{\omega} - 2 N g s_x x
    + N f_\text{Ising}(\beta, \heff, J),
\end{multline*}
where $f_\text{Ising}=F_\text{Ising}/N$ is the free energy per particle for the atoms as calculated in Eq. (\ref{eq:IsingFreeEnergy}). Note that the free energy includes a term representing the thermal distribution of the cavity-photons as well as the contribution depending on the field amplitude. The values of the mean-fields can now be obtained by minimizing the free energy. A short calculation reveals
  \numparts
\begin{align}
    \frac{1}{N}\frac{\partial F}{\partial s_x} &= -\frac{2 g^2 s_x}{\omega} - 2 g x \label{eq:FreeEnergyMinimumA} \\
    \frac{1}{N}\frac{\partial F}{\partial x} &= -2 g s_x + \frac{\partial f_{Ising}}{\partial x}, \label{eq:FreeEnergyMinimumB}
\end{align}
  \endnumparts
This shows that $s_x = -x\omega/g$ and that the Dicke-order parameter $x$ should be found by minimizing $\omega x^2 + f_\text{Ising}$ with respect to $x$. We will return to this minimization problem below.\par

In the mean-field description the interpretation of the physical properties of the system becomes clear and unambiguous. As an example, with the mean-field theory we can obtain expressions for various correlation functions for the atomic variables from the large amount of theory already present on the transverse Ising chain since in thermodynamic equilibrium the strongly coupled system is effectively identical to a rotated Ising chain.

It is reassuring, but hardly surprising, that the mean field result can be recovered from the coherent state integral for the thermodynamic limit $N\to\infty$ result. By Laplace's (saddle point) method, one can replace the integral over coherent state amplitudes by discrete contributions from the location of the maximum of the exponential with respect to $z$ in (\ref{eq:CoherentStatePartitionFunction}). It is easily shown that this maximization coincides with the mean field, identified by minimization of the free energy and by picking one of the maximizers in (\ref{eq:CoherentStatePartitionFunction}) to represent a symmetry-broken physical state of the system.

\section{Phase diagram and analysis of the free energy} \label{sec:Phasediagram}
\begin{figure*}
 \begin{center}
 \includegraphics[width=0.49\textwidth]{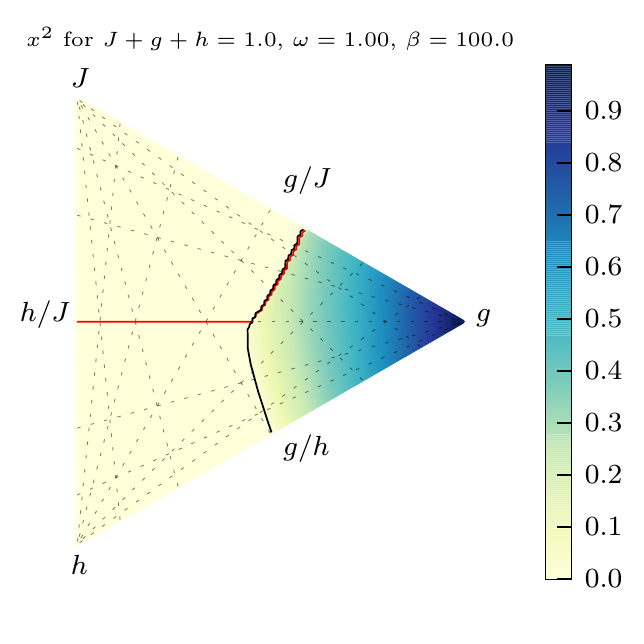}
 \includegraphics[width=0.49\textwidth]{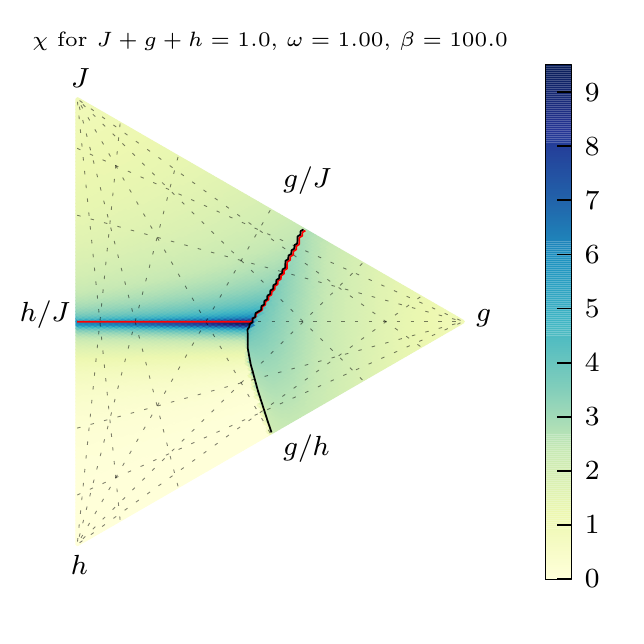}
  \caption{(Colour online) Observables for the Dicke-Ising model with $\beta = 100$, $\omega = 1$ and $g + J + h = 1$. Left: order parameter $x^2$ and right: susceptibility along the $z$-axis $\chi = \partial m/\partial h$. All dotted lines converging to a corner correspond to a fixed ratio of the parameters indicated on the edge opposing the corner. The black solid line is where the order parameter $x$ becomes non-zero, corresponding to the Dicke phase transition for $J = 0$, and the red solid line is where the $\heff / J = 1$,  corresponding for $g = 0$ to the Ising phase transition. }
\label{fig:observables1}
\end{center}
\end{figure*}

We now turn to the problem of finding the minimum of the free energy (\ref{eq:FreeEnergyMinimumA}),(\ref{eq:FreeEnergyMinimumB}) along with several important observables like the magnitude of the oscillator mean field, the spin magnetization and the susceptibility $\chi = \partial m_z/\partial h$.

\begin{figure*}
 \begin{center}
 \includegraphics[width=0.49\textwidth]{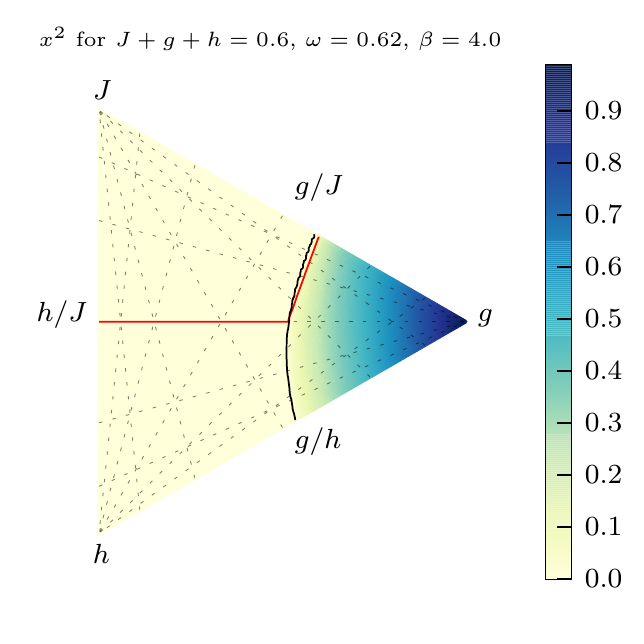}
 \includegraphics[width=0.49\textwidth]{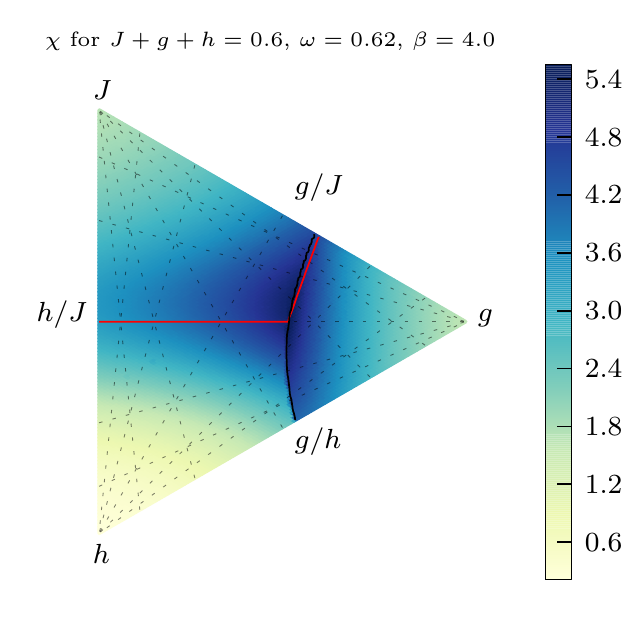}
  \caption{(Colour online) Observables for the Dicke-Ising model with $\beta = 4$, $\omega = 0.62$ and $g + J + h = 0.6$. Left: order parameter $x^2$ and right: susceptibility along the $z$-axis $\chi = \partial m/\partial h$. All dotted lines converging to a corner correspond to a fixed ratio of the parameters indicated on the edge opposing the corner. The black solid line is where the order parameter $x$ becomes non-zero, corresponding to the Dicke phase transition for $J = 0$, and the red solid line is where the $\heff / J = 1$,  corresponding for $g = 0$ to the Ising phase transition. }
\label{fig:observables2}
\end{center}
\end{figure*}

For fixed $\beta$ and $\omega$, the system is controlled by three parameters $h,\ J$ and $g$. To illustrate the phase transitions in the system, we introduce a convenient way to plot different quantities as function of these variables in Fig. \ref{fig:observables1} and \ref{fig:observables2}. In each plot, the sum $h+J+g$ is fixed, and the corners of the triangles shown correspond to each of the three quantities acquiring the maximum value while the others vanish. The straight dotted lines converging to the corners of the triangles correspond to definite values of the ratio between the two quantities indicated on the edges of the triangles. These plots can be thought of as slices of the three-dimensional simplex defined by $h + J + g + \omega = \epsilon/\beta$ where $\epsilon$ should be interpreted as the system energy-scale. In this coordinate representation, we show with colour coding the value of different interesting quantities.

Of particular interest in Fig. \ref{fig:observables1} is the oscillator field strength, represented by $x^2$, and the susceptibility $\chi$ which are shown for the case where $\omega = 1$, $\beta = 100$ and $h + J + g = 1$ for Fig. \ref{fig:observables1} and $\omega = 0.25$, $\beta = 4.0$ and $h + J + g = 1$ for Fig. \ref{fig:observables2} where $h$, $J$ and $g$ are positive.

The edge of the triangle between $g$ and $h$ (i.e., with vanishing $J$) corresponds to the usual Dicke-model, while the edge between $h$ and $J$ (i.e., with vanishing $g$) corresponds to the usual Ising-model with the critical point at $h/J = 1$, showing up clearly as a signature in the variation of the susceptibility $\chi$. We observe that this signature is present also for finite Dicke coupling parameter in the plot. The black curve in each plot shows where the Dicke phase transition occurs. In both Fig. \ref{fig:observables1} and Fig. \ref{fig:observables2} signatures of both first- and second-order phase transitions can be seen. Approximately below the line $h/J = 1$ the second order transition can be identified by the smooth increase in $x^2$ whereas above $h/J = 1$ one can discern a discontinuous jump in the order parameter.\par

The most significant difference between the cases presented in Figs. \ref{fig:observables1} and \ref{fig:observables2} is the susceptibility $\chi$. For moderately low temperatures, $\beta = 4$ in Fig. \ref{fig:observables2}, the signature of the Ising quantum phase transition is still clearly present, whereas for very low temperatures, $\beta = 100$ Fig. \ref{fig:observables1}, the Ising phase transition becomes almost completely suppressed in the Dicke regime. Indeed, far into the Dicke regime (towards the right vertex in the triangles) the spin interactions do not appear to play any significant role.

Looking at the black and red solid lines in Fig. \ref{fig:observables1} one might be tempted to conclude that in the super-radiant phase we always have $\heff/J > 1$ (i.e. the red line does not penetrate into the area to the right of the black line). A close look at Fig. \ref{fig:observables2} will, however, reveal that this is not always the case. Indeed, for $h + J + g = 0.6$, $\omega = 0.62$ and $\beta = 4$ there is a small part of the parameter-space where $x > 0$ and $\heff < J$. The physical reason why the Dicke phase transition almost, but not quite, suppresses the Ising transition remains to be understood. Formally, it occurs because the free energy has a minimum giving an effective magnetic field such that $\heff/J > 1$. The Ising critical point is therefore simply skipped in these cases and only an amputated signature of the Ising phase transition is present in cases such as shown in Fig. \ref{fig:observables1}.

To investigate this phase transition in more detail, let us further consider the minimization of the free energy. The order-parameter $x$ enters the mean field Ising-term via the effective magnetic field $(\heff/J)^2 = \tilde h^2 + 4 g^2 x^2 / J^2$. If we therefore introduce a rescaled order-parameter $\tilde x = 2 g x/J$ and a rescaled mode frequency $\tilde\omega = \omega J/4g^2$ we can write the free energy as a function of a few dimensionless quantities
\begin{widetext}
\begin{align}
 \frac{F(\tilde x)}{N J} = -\int_{-\pi}^\pi \frac{\d k}{2\pi \tilde\beta} \log(2\cosh(\tilde\beta (1 + (\tilde h^2 + \tilde x^2) - 2 (\tilde h^2 + \tilde x^2)^{1/2} \cos(k))^{1/2})) \nonumber \\
  + \tilde\omega \tilde x^2 + C, \label{eq:ReducedFreeEnergy}
\end{align}
\end{widetext}
where $C$ is a constant independent of $\tilde x$.

\begin{figure}[htb]
 \includegraphics[width=\columnwidth]{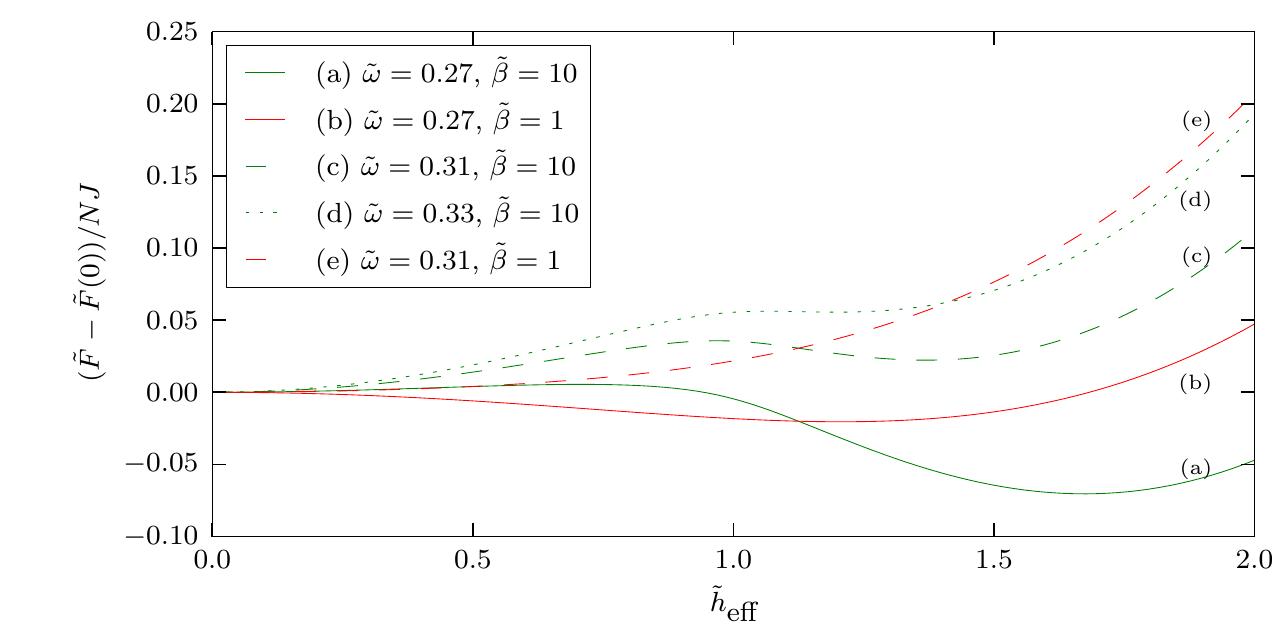}
 \caption{(Colour online) Examples of various functional shapes of $(\tilde F - \tilde F(0))/NJ$ as a function of $\heff$. The different line styles indicate varying $\tilde\omega$ and the two colours indicates $\tilde\beta = 1$ and $\tilde\beta = 10$ respectively.}
 \label{fig:FreeEnergyExamples}
\end{figure}

In order to understand the structure of the phase diagram it is necessary to investigate how the integral changes as a function of $\tilde\beta = \beta/J$ and $\tilde h$ compared to $\tilde\omega \tilde x^2$. If we choose the variable $\tildeheff = \heff/J = \sqrt{h^2 + 4g^2 x^2}/J$ as the independent variable instead of $\tilde x$, the integrand in (\ref{eq:ReducedFreeEnergy}) only depends on $\tilde\beta$ and the new variable $\tildeheff$, while $\tildeheff \geq \tilde h$ imposes a boundary condition on the minimization with respect to $\tildeheff$. To avoid confusion, we will consider $F$ a function of $\tilde x$ and use the symbol $\tilde F$ to denote the dependence on $\tildeheff$. The system is in the super-radiant phase whenever the minimum in $\tilde F$ occurs for $\tildeheff > \tilde h$ which implies $x \neq 0$. Examples of $\tilde F$ for representative values of $\tilde\omega$ and $\tilde\beta$ can be seen in figure \ref{fig:FreeEnergyExamples}. By a numerical investigation it is quickly revealed, that $\tilde F$ has at most a single local minimum (e.g. the curves (a), (b), (c) and (d) in figure \ref{fig:FreeEnergyExamples}) at $\tildeheff \neq 0$ or no local minimum (curve (e) in figure \ref{fig:FreeEnergyExamples}). The existence and location of the minimum are thus solely determined by $\tilde\beta$ and $\tilde\omega$.

This implies that when keeping $\tilde\omega$ and $\tilde\beta$ fixed the minimum of the free energy is either at $\tildeheff = \tilde h$ or at the local minimum of $\tilde F$. If we imagine tuning $\tilde h$ from high values towards low values (i.e, setting the boundary condition $\tildeheff \geq \tilde h$ at different locations, for example along curve (c) in figure \ref{fig:FreeEnergyExamples}) the system will pass a second order phase-transition when $\tilde h$ passes the local minimum of $\tilde F$. In the case that the local minimum is not the global minimum there will be a $\tilde h > 0$ where $\tilde F$ goes below its value at the local minimum implying that when one further lowers $\tilde h$ the system will undergo a first-order phase transition into the normal state again. By the same reasoning if $\tilde F$ has a single global minimum (curve (a) and (b) in figure \ref{fig:FreeEnergyExamples}) there can only be a second order phase transition when tuning $\tilde h$.

The second-order phase transitions can be investigated in further detail using Gintzburg-Landau theory: In the neighbourhood of the second-order phase transition the order parameter $\tilde x$ will always be small so we can expand the free energy $F$ as a polynomial in $\tilde x$ around $\tilde x = 0$:
\begin{align}
 \frac{F(\tilde x)}{NJ} \approx C + I_0(\tilde\beta, \tilde h) + (\tilde\omega + I_2(\tilde\beta, \tilde h) ) \tilde x^2 + I_4(\tilde\beta, \tilde h) \tilde x^4,
\end{align}
where $I_n$ is the $n$'th term in the Taylor expansion of the integral (\ref{eq:ReducedFreeEnergy}) with respect to $\tilde x$. The standard argument from Gintzburg-Landau theory is now that this fourth-order polynomial has a non-zero minimum when $\tilde\omega + I_2(\tilde\beta, \tilde h)$ is negative. The second order phase-transition therefore occurs when $I_2(\tilde\beta, \tilde h) = -\tilde\omega$. By numerical investigation one finds that $-I_2$ is bounded by approximately $0.3356$ implying that for $\tilde\omega > 0.3356$ no phase transition can occur.

The first-order phase transition grows out of the second-order phase transition so there will be a region where the first-order jump in the order parameter is small. In that case we can still use Ginzburg-Landau theory and in particular we can find the point where the second-order transition changes to a first-order transition, i.e. when a local minimum in $F$ changes from purely local to truly global. Again we analyze the polynomial expansion and one can show \cite{plischke_equilibrium_1994} that one needs to solve the system of equations $I_4(\tilde\beta, \tilde h) = 0$ and $I_2(\tilde\beta, \tilde h) + \tilde\omega = 0$  to obtain the point where the phase transition changes nature. By investigating the functional form of $I_4$ it turns out that there is a minimal $\beta_c$ below which the first order phase-transition cannot occur. This value can be calculated numerically and is approximately $\tilde \beta_c \approx 1.1430$.

\begin{figure}[htb]
 \includegraphics[width=\columnwidth]{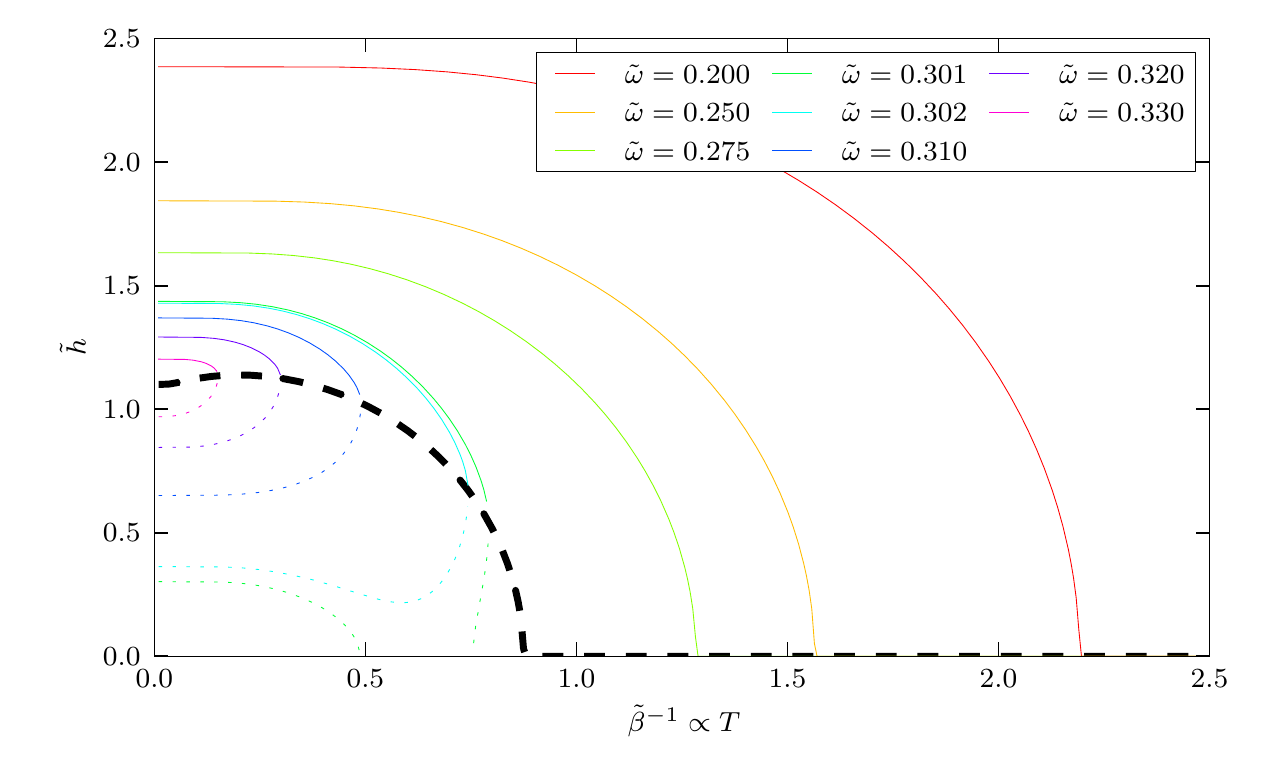}
 \caption{(Colour online) Phase diagram for the Dicke-Ising model. Each curve shows the location of the phase transition for the specified parameters. If the curve is full the transition is second order, whereas a dotted curve represents a first order transition. The dashed black curve represents the general boundary $I_4(\tilde\beta,\tilde h) = 0$ where the transition changes from second to first order along each of the coloured curves.}
 \label{fig:Phasediagram}
\end{figure}

This identifies where the second-order transition changes to a first-order transition. To determine the first-order transition boundary for finite jumps in the order parameter, however, it is necessary to deal with the free energy $F$ to all orders. Numerically it is not difficult to investigate for which value of $\tilde h$ the value of $\tilde F$ coincides with the value at the local minimum as described above. All this information has been combined into figure \ref{fig:Phasediagram} where the phase-boundaries for various values of $\tilde\omega$ have been indicated. The coloured dotted lines represent first order transitions whereas the solid lines indicate second-order transitions. The black dashed curve indicates where $I_4(\tilde\beta, \tilde h) = 0$ and its intercept with the curves $I_2(\tilde\beta, \tilde h) + \tilde\omega = 0$ indicates where the phase transition changes type between first- and second-order transitions.

\section{Using a phase transition for a high precision measurement}
A first order phase transition is interesting for many different reasons and here we consider its use as a measurement tool. Indeed, the standard description of a first order phase transition includes a discontinuous jump in the order parameter, and it is a relevant question, how precisely an experiment can locate the position of this discontinuous jump. The size of the jump-discontinuity usually scales linearly with the number of particles, while the width of the transition region often scales with an inverse power of this number, and under that assumption we shall present a simple model for the metrological sensitivity of the system. Since the phase transition occurs for rather non-trivial combinations of the temperature and the interaction parameters of the models, by changing some of these parameters in a controllable way, one may be able to select parameter ranges with particularly high sensitivity of the phase transition point to the value of the quantity being probed.

\begin{figure}
 \includegraphics[width=\columnwidth]{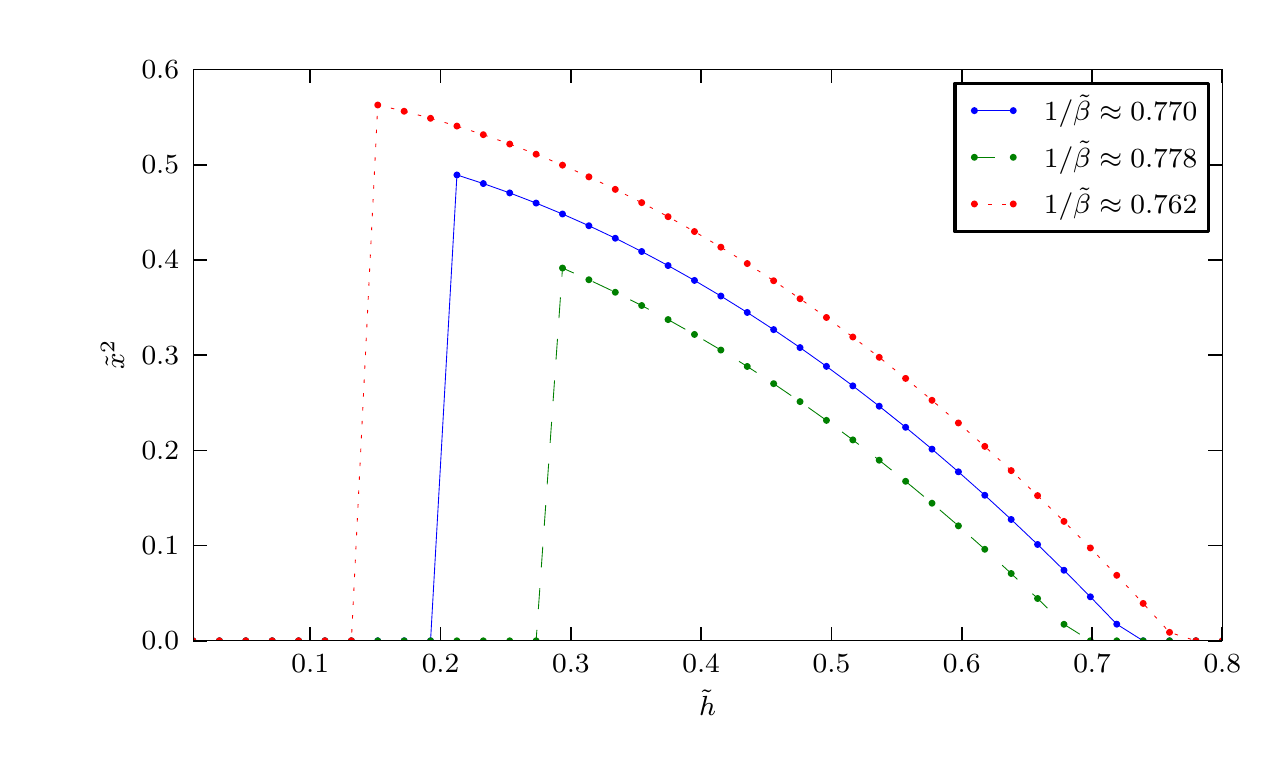}
 \caption{(Colour online) The order-parameter $\tilde x^2$ as a function of $\tilde h$ and $\tilde \beta$ for the lower-right part of the curve for $\tilde\omega = 0.301$ in Fig. \ref{fig:Phasediagram} where the slope of the phase transition line is large. For the blue curve $1/\tilde\beta = 0.77$, for the green curve the temperature is 1\% higher and for the red curve the temperature is 1\% lower.} \label{fig:OrderparameterCut}
\end{figure}

We will consider a measurement strategy where the system is probed at some range of values of some control parameter, e.g. a bias magnetic field. Fig. \ref{fig:OrderparameterCut} shows how the cavity field order parameter varies as a function of $\tilde h$ for three different temperatures. To produce this plot, we have selected values of $\tilde{\omega}$ and thus an area of the phase diagram  where the critical magnetic field depends strongly on the temperature, cf. the steepness of the dashed curves in Fig. \ref{fig:Phasediagram}. We expect that it will be possible to determine the critical value of the magnetic field with high precision, and since in this case a variation of the temperature of 1\% changes the critical magnetic field by approximately 35\%, the measurement of the critical field yields a very sensitive temperature measurement within the appropriate range of values $\tilde\beta \approx 0.77 \pm 1\%$. Sensitivity in, e.g., a lower temperature range is obtained if we chose a higher value of $\tilde{\omega}$ and scan a different range of values of the magnetic field.

The order parameter presented here is the intra cavity field intensity. We imagine that the cavity leaks photons at a sufficiently low rate not to significantly disturb the thermodynamic steady state of the system, and herewith, detection of the intensity of the emitted light is a direct probe of the cavity field order parameter. We assume that the inverse temperature $\tilde\beta$ is known to be close to some reference value, and we can then estimate the difference $\delta\tilde\beta$ by the best unbiased linear estimator as described in Appendix \ref{sec:BLUE}. This estimator is given by
\begin{align*}
 \hat{\delta\tilde\beta}( \{n_i\} ) = \frac{1}{N} \sum_i \frac{ \mu'(\tilde h_i) }{\sigma^2(\tilde h_i) } (n_i - \mu(\tilde h_i) )
\end{align*}
where $n_i$ is the detected number of photons in a given time while the controllable effective bias field $\tilde{h}$ attains the value $\tilde h_i$. $\mu(\tilde h_i)$ is the expectation value of the photon number and $\sigma^2(\tilde h_i)$ is the photon number variance. $\mu'(\tilde h)$ denotes the derivative of the expected photon number with respect to changes in inverse temperature $\tilde{\beta}$, and the expression applies within a narrow range where a linear variation of the expected photon number with $\tilde\beta$ is valid.

In the limit of high bias field resolution, the sum in the estimator can be converted into an integral, and one can determine the variance of the estimate: $\Var( \hat{\delta\tilde\beta}(n(\tilde h)) = 1/\int \mu'(\tilde h)^2 / \sigma^2(\tilde h) \d \tilde h$, see details in the appendix.

So far the arguments have been of a general nature. Let us now assume the Dicke-Ising model, in which the photon number distribution is well described as a thermal state below and a displaced thermal state above  the Dicke phase transition. The first and second moments of such distributions can be calculated using, e.g. the positive P-representation for the thermal state,
\begin{align*}
 \mu(\tilde h, \tilde\beta) &= \braket{n} = N x^2 + \bar n \\
 \sigma^2(\tilde h, \tilde\beta) &= \Var n = N x^2 (1 + 2\bar n) + \bar n + \bar n^2
\end{align*}
Recall that the order parameter $x^2$ is a function of the system parameters $\tilde h$, $\tilde\beta$ and $\tilde\omega$ and in the thermodynamic limit it has a discontinuous jump at the dashed lines shown in Fig. \ref{fig:Phasediagram}. For a finite system, however, the phase transition constitutes a smooth curve with a fast increase of the order parameter. The width of this region is not easy to determine but finite size effects in phase transitions tend to smoothen phase transitions leading to a decreasing width as $N$ increases. Indeed, \cite{imry_finite-size_1980} finds that, in general, the width scales as $1/N$, but for the sake of generality we assume a scaling $N^{-\gamma}$, $\gamma > 0$.

Since both $\braket{n}$ and $\Var n$ are proportional to $N x^2$ both $\mu(h)$, $\mu'(h)$ and $\sigma^2(h)$ will carry a signature of this power law. To be explicit assume that $N \tilde x^2(\tilde h, \tilde\beta) = N \tanh(N^\gamma (\tilde h - \tilde h_c(\tilde\beta))$ where $\tilde h_c$ is the critical value of $\tilde h_c$ as a function of $\tilde\beta$. Then $\mu(\tilde h) \propto N$, $\mu'(\tilde h) = (\partial_{\tilde\beta} \mu)(\tilde h, \tilde \beta_0) \propto N^{1+\gamma} \tilde h_c'(\tilde\beta_0)$ and $\sigma^2(\tilde h) \propto N$. The function $\mu'$ only has support in a region of width $1 / N^\gamma$ near $\tilde h_0 \equiv \tilde h_c(\tilde\beta_0)$. The variance of the estimate $\hat{\delta\tilde\beta}$ then scales as $\Var{\hat{\delta\tilde\beta}} \approx \left( (\mu'(h_0)^2/\sigma^2(h_0)) N^{-\gamma}\right)^{-1} \propto (\tilde h_c'(\tilde\beta_0)^2 N^{2+2\gamma}/N)^{-1} N^{\gamma}$, i.e. 
\begin{align*}
 \Var{\hat{\delta\tilde\beta}} \propto \frac{1}{\tilde h_c'(\tilde\beta_0)^2  N^{1 + \gamma}}
\end{align*}

This is our main result of this section, showing that the sensitivity is better than the "standard limit" where one expects $\Var \tilde\beta \sim 1/N$, and depending on the character of the finite size effects (the power $\gamma$), it is potentially also better than the Heisenberg detection limit. With the result from \cite{imry_finite-size_1980}, $\gamma = 1$ the accuracy is actually at the Heisenberg limit. Note that the above argument is quite general and applies to any first order transition with an intensive order parameter.\par

The term $\tilde h_c'$ is included in our expression in order to show explicitly that the sensitivity depends on the curve of critical points in the phase diagram of the system. From Fig. \ref{fig:Phasediagram} we see that $\tilde h_c'$ can be chosen large for arbitrarily small temperatures by tuning $\tilde\omega$. The large value of $\tilde h_c'$, however, comes at a cost: The slope of $h_c$ is highest near the thick dashed curve, which is also where the first order transition has small amplitude and changes to a second order transition. In a concrete implementation, the values of $\tilde\omega$ and the range of effective magnetic fields need to be chosen with care to reflect the actual scaling $N^{-\gamma}$ and the size of the jump discontinuity.

With an adaptive measurement scheme, we imagine that the number of iterations with different $\tilde h_i$  for a reliable detection of the critical value of the $\tilde h$-parameter can be optimized. It is  clear that a more detailed investigation is necessary in order to quantify the accuracy and scaling of resources of such measurements. Indeed, the specific power law  $1/N^\gamma$ for the transition width is only a convenient Ansatz, and a non-mean field calculation on a finite system will be needed in order to investigate the approach towards the thermodynamic limit in more detail. Furthermore, the critical properties and the long range correlations of the system may possibly lead to even better estimates by use of the recent techniques of quantum non-linear parameter estimation \cite{choi_bose-einstein_2008,boixo_generalized_2007}. These issues we shall defer to a later publication.

\section{Conclusion} \label{sec:Conclusion}
We have investigated the thermodynamic properties of a Dicke-Ising model incorporating both the quantum transverse Ising model and the Dicke model as special limiting cases. We have derived expressions for the free energy using a coherent state integral similar to \cite{wang_phase_1973} but also using a mean field theory with a clearer interpretation for the field statistics. The combined model exhibits a first order phase transition which is not present in either of the two separate models. By a simple numerical search the free energy minimum can be identified and the value of the Dicke mean field and the magnetic susceptibility can be determined as functions of all physical parameters of the model, cf. Figs. \ref{fig:observables1} and \ref{fig:observables2}.

Using the free energy and Ginzburg-Landau theory we also investigated the complete phase diagram as shown in Fig. \ref{fig:Phasediagram}. The Dicke phase transition occurs also for moderate inter-particle interactions and the Ising phase transition is also well preserved for weak and moderate light-matter couplings. In a small area of the parameter space both phase transitions coexist closely together, but for a stronger Dicke model interaction the resulting mean field puts the system in a regime without any observable Ising phase transition.

The Dicke-Ising model constitutes an interesting mix of second-, first- and infinite order phase transitions. The interplay of these phase transitions and a complete description beyond the mean field approximation of the fundamental excitations at the critical points  would be and interesting continuation along the lines of this work. In addition to its fundamental theoretical interest, the first order phase transitions provides a tool for precise measurements of, e.g. the magnetic bias field or of the temperature. We have presented a simple estimate of the accuracy of such a measurement device showing that the variance scales as $1/N^{1+\gamma}$ which is better than the standard limit $1/N$ for independent measurements on $N$ particles.

This work was supported by the European Union Integrated Project AQUTE.

\appendix

\section{Best unbiased linear estimator} \label{sec:BLUE}
We will consider the best unbiased linear estimator in a situation where an experimenter performs a sequence of measurements where she scans a parameter $x$ (e.g. a bias magnetic field) in order to uncover another, unknown, parameter $q$ (e.g. the location of a critical point). The experimenter measures a discrete stochastic variable $n$ (e.g. photon number) which has probability distribution $p(n; x, q)$. In the measurement the experimenter thus collects, for a fixed $q$, the values $n_i$ corresponding to selected $x_i$ .\par

In the following we assume that the unknown parameter $q$ is close to a reference value, which, without loss of generality, we take to be zero. For $q \ll 1$ we can then expand the moments of $p$ in a Taylor expansion such that
\begin{align*}
 \mean{n_i} &= \mu(x, q) \approx \mu(x) + q \mu'(x) \\
 \Var(n_i) &= \sigma^2(x, q) \approx \sigma^2(x)
\end{align*}
where we have expanded the mean $\mu(x, q)$ to first order (and $\mu(x) \equiv \mu(x, q)$, $\mu'(x) \equiv (\partial_q \mu)(x, 0)$) and the variance $\sigma^2(x, q)$ to zeroth order in $q$.

A linear estimator is of the form
\begin{align*}
 \hat q( \{ n_i \} ) = \sum_i g_i n_i + c,
\end{align*}
where $\{g_i\}$ are weighting coefficients, $\{n_i\}$ are the observed values of $n_i$ corresponding to the chosen values $x_i$ and $c$ is a constant. To find an unbiased estimator we require $\mean{\hat q( \{ n_i \} )} \propto q$ which implies $c = -\sum_i g_i \mu^0(x_i)$. To find the best linear estimator we optimize the signal-to-noise ratio $\mean{q( \{ n_i \} )}^2/\Var(q( \{ n_i \} ))$ with respect to the vector $g_i$. To second order in $q$ the signal-to-noise ratio is $q^2 (\sum_i g_i \mu'(x_i) )^2/\sum_i g_i^2 \sigma^2(x_i)$. The minimum of the signal-to-noise ratio is obtained when $g_i \propto \mu'(x_i)/\sigma^2(x_i)$.

The constant of proportionality should then be chosen such that $q = \mean{q( \{ n_i \} )} = \sum_i g_i (\mean{n_i} - \mu(x_i) )$ which gives the condition $\sum_i g_i \mu'(x_i) = 1$ and the normalization constant is given by $A^{-1} = \sum_i (\mu'_i)^2/\sigma^2_i$. With this normalization the variance of the estimate $q$ is
\begin{align}
  \label{eq:BLUEVariance}
 \Var(\hat q(\{n_i\} )) = \left(\sum_i \frac{ (\mu'(x_i) )^2 }{\sigma^2(x_i) } \right)^{-1}
\end{align}

\bibliography{litteratur.bib}

\end{document}